\documentclass[twocolumn,showpacs,preprintnumbers,amsmath,amssymb]{revtex4}

\usepackage{graphicx}
\usepackage{dcolumn}
\usepackage{bm}

\begin{document}

\preprint{}

\title{Controlling the interactions of a few cold Rb Rydberg atoms \\ by radio-frequency-assisted F\"orster resonances}
\author{D.~B.~Tretyakov$^1$}
\author{V.~M.~Entin$^1$}
\author{E.~A.~Yakshina$^{1, 2, 3}$}
\author{I.~I.~Beterov$^{1, 2}$}
\author{C.~Andreeva$^{4, 5}$}
\author{I.~I.~Ryabtsev$^{1, 2, 3}$}
  \email{ryabtsev@isp.nsc.ru}
\affiliation{$^1$Rzhanov Institute of Semiconductor Physics SB RAS, 630090 Novosibirsk, Russia }
\affiliation{$^2$Novosibirsk State University, 630090 Novosibirsk, Russia}
\affiliation{$^3$Russian Quantum Center, Skolkovo, Moscow Region, 143025, Russia}
\affiliation{$^4$University of Latvia, LV-1002 Riga, Latvia }
\affiliation{$^5$Institute of Electronics, Bulgarian Academy of Sciences, Sofia 1784, Bulgaria }
\date{31 October 2014}

\begin{abstract}

Long-range interactions between cold Rydberg atoms, which are used in many important applications, can be enhanced using F\"orster resonances between collective many-body states controlled by an external electric field. Here we report on the first experimental observation of highly-resolved radio-frequency-assisted F\"orster resonances in a few cold Rb Rydberg atoms. We also observed radio-frequency-induced F\"orster resonances which cannot be tuned by a dc electric field. They imply an efficient transition from van der Waals to resonant dipole-dipole interaction due to Floquet sidebands of Rydberg levels appearing in the rf-field. This method can be applied to enhance the interactions of almost arbitrary Rydberg atoms with large principal quantum numbers.

\end{abstract}

 \maketitle

Long-range interactions between highly-excited Rydberg atoms are being investigated for several important applications like neutral-atom quantum computing [1], quantum simulations [2], phase transitions in cold Rydberg gases [3], or nonlinear optics with single photons [4]. Depending on the particular Rydberg states, these are van der Waals (vdW) or dipole-dipole (DD) interactions with different dependences on interatomic distance \textit{R} ($R^{-6} $ and $R^{-3} $, correspondingly). 

Atoms in an identical \textit{nL} Rydberg state generally interact via vdW, which is much weaker than DD at long distances (longer than the Rydberg atom size that scales as $n^{2} $). To make identical atoms interact via DD, the Rydberg state should be tuned exactly midway between two other Rydberg states of the opposite parity to induce a F\"orster resonance [5]. It can be tuned using the Stark effect in a dc electric field. This method, however, works only for a limited number of Rydberg states. For example, in Rb atoms, narrow Stark-tuned F\"orster resonances between neighboring Rydberg states can be obtained for $nP_{3/2} $ states with $n\le 38$ [6-8], $nD_{3/2} $ states with $n\ge 40$, and $nD_{5/2} $ states with $n\ge 43$ [9-11].  

Other methods are thus required to control the interactions and tune vdW to DD for arbitrary Rydberg states. One such method was demonstrated in Refs. [9], where the energies of Rb Rydberg states were adjusted by an ac Stark shift in a strong nonresonant microwave field at 28.5 GHz or in a nearly resonant to a Rydberg transition microwave field at 1.356 GHz. Another method is to apply a resonant microwave field that drives a transition between Rydberg states of the opposite parity and mixes them up [12-14].

It is also possible to control Rydberg interactions by microwave-assisted F\"orster resonances [15-19], when one or several microwave photons compensate for the energy defect and induce transitions between the initial and final many-body collective states of the F\"orster resonance. Microwave or even radio-frequency (rf) photons can  provide the tunability of F\"orster resonances in a wide range. 

In this Rapid Communication we demonstrate that rf electric fields can be used to induce "inaccessible" F\"orster resonances, which cannot be tuned by a dc electric field, and that it leads to an efficient transition from vdW to DD interaction for a few cold Rb Rydberg atoms. A new point compared to Refs.~[9,12-14] is that the rf-field drives the transitions not between Rydberg states of a single atom (typical frequencies 10-100 GHz) but between nearly degenerate collective states of the quasi-molecule formed by the interacting Rydberg atoms (typical frequencies 10-100 MHz).  

The process under study is the F\"orster resonant energy transfer ${\rm Rb}(nP_{3/2} )+{\rm Rb}(nP_{3/2} )\to {\rm Rb}(nS_{1/2} )+{\rm Rb}([n+1]S_{1/2} )$ due to the dipole-dipole interaction of two or more cold Rb Rydberg atoms in a small laser excitation volume [8,20,21]. The energy detuning of this resonance $\hbar \Delta =E(nS_{1/2})+E([n+1]S_{1/2}) -2E(nP_{3/2})$ is controlled by a weak dc electric field. It can be tuned to zero for Rydberg states with $n\le 38$ as shown in Fig.~1(a) for the $37P_{3/2} $ state, while for states with $n\ge 39$ the dc electric field increases $\Delta$ and the resonance could be induced only by the rf-field [see Fig.~2(a) for the $39P_{3/2} $ state]. 

Experiments were performed with cold $^{85}$Rb atoms in a magneto-optical trap. The excitation to the \textit{nP}$_{3/2}$(\textbar M$_J$\textbar =1/2) Rydberg state is realized via three-photon transition $5S_{1/2} \to 5P_{3/2} \to 6S_{1/2} \to nP_{3/2} $ by means of three cw lasers modulated to form 2~$\mu $s exciting pulses at a repetition rate of 5~kHz [22]. The small Rydberg excitation volume of 30-40 $\mu $m size is formed using crossed-beam geometry [20]. 

Our experiment provides atom-number-resolved measurement of the signals obtained from $N=1-5$ of the detected Rydberg atoms with a detection efficiency of 65\% [8]. It is based on selective field ionization (SFI) detector with channel electron multiplier and post-selection technique [23]. The normalized \textit{N}-atom signals $S_{N}$ are the average fractions of atoms that have undergone a transition to the final \textit{nS} state. In fact, the signals measured in our experiment correspond to the detection of \textit{N}-body collective states, which are used in theoretical calculations [8,21]. Compared to the previous works on rf-assisted F\"orster resonances [17,18] we deal with a few Rydberg atoms in an identical $nP_{3/2} $ state, interacting in a single small excitation volume.  

We use Stark-switching technique [7,11] to switch the interactions on and off as depicted in Fig.~1(b). The laser excitation occurs for 2~$\mu $s at a fixed electric field of 5.6~V/cm. Then the field decreases to a lower value near the resonant electric field (1.79~V/cm for the 37\textit{P}$_{3/2}$ state), which acts for 3~$\mu $s until the field increases back to 5.6~V/cm. The lower electric field is slowly scanned across the F\"orster resonance and SFI signals are accumulated for 10$^{3}$-10$^{4}$ laser pulses. An rf pulse with variable amplitude (0-300~mV) and frequency (10-100 MHz) is superimposed on the lower dc field. 

Figures~1(c)-1(f) present the spectra $S_{N} $ of the F\"orster resonance ${\rm Rb}(37P_{3/2} )+{\rm Rb}(37P_{3/2} )\to {\rm Rb}(37S_{1/2} )+{\rm Rb}(38S_{1/2} )$ in a 15 MHz rf-field of various amplitudes recorded for \textit{N}=2-5 detected Rydberg atoms as a function of the dc electric field. Without the rf-field in Fig.~1(c), the single narrow peak at 1.79~V/cm is the "true" Stark-tuned F\"orster resonance, whose amplitude and width grow with \textit{N} according to theory [8,21]. The spectrum $S_{2} $ corresponds to DD interaction of just two Rydberg atoms in the interaction volume [8]. Its width of 16 mV/cm is equivalent to the frequency width of 1.9~MHz.

\begin{figure}
\includegraphics[scale=0.47]{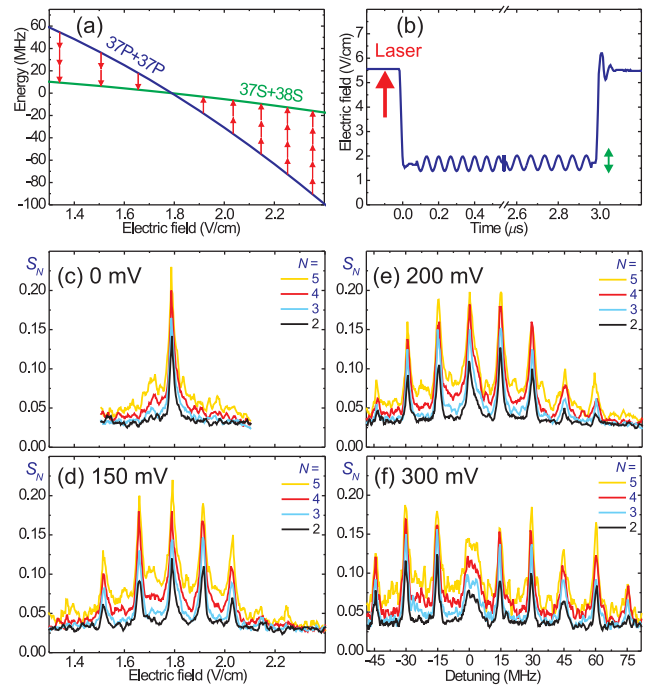}
\caption{\label{Fig1} (color online). (a) Energy levels of the initial 37\textit{P}+37\textit{P} and final 37\textit{S}+38\textit{S} collective states of two interacting Rb Rydberg atoms in an electric field. The arrows indicate rf-induced F\"orster resonances ${\rm Rb}(37P_{3/2} )+{\rm Rb}(37P_{3/2} )\to {\rm Rb}(37S)+{\rm Rb}(38S)$ of various orders at rf-frequency 15~MHz. (b) Time dependence of the electric field. (c) Single F\"orster resonance without rf-field for \textit{N}=2-5  detected Rydberg atoms as a function of the electric field. (d) RF-assisted F\"orster resonances at 150 mV rf-amplitude, for the same scale. (e)-(f) RF-assisted F\"orster resonances as a function of frequency detuning at 200 and 300~mV rf-amplitudes, respectively.}
\end{figure}

Application of a 150~mV rf-field [Fig.1(d)] induces additional F\"orster resonances, which are  rf-assisted resonances of various orders, as shown in the scheme in Fig.~1(a) for the energy levels of the initial 37\textit{P}+37\textit{P} and final 37\textit{S}+38\textit{S} collective states in the dc electric field. The arrows indicate rf-induced F\"orster resonances of different orders. As \textit{N} increases, the resonance amplitudes grow and the rf-assisted resonances become more pronounced in Fig.~1(d)  due to increase in the total DD interaction energy. 

The frequency interval between the peaks in Fig.~1(d) corresponds exactly to 15~MHz, taking into account the known polarizabilities of these Rydberg states [20]. Figures~1(e)-(f) present such spectra as a function of frequency detuning for 200 and 300 mV rf-amplitudes. The observed high-order resonances have almost the same amplitude and width as the low-order ones if the rf-amplitude is large enough. This means that the vdW undergoes a transition to DD with high efficiency, reaching 50-100\%. By changing the rf-frequency we were able to control the positions of the peaks with high precision.

Now we turn to the "inaccessible" F\"orster resonances, which cannot be tuned by dc electric field. An example is the F\"orster resonance ${\rm Rb}(39P_{3/2} )+{\rm Rb}(39P_{3/2} )\to {\rm Rb}(39S_{1/2} )+{\rm Rb}(40S_{1/2} )$ whose collective energy levels in the dc electric field are shown in Fig.~2(a). The dc field alone increases the energy detuning $\Delta $ and makes the interaction weaker. However, our experience with the F\"orster resonance for the 37\textit{P}$_{3/2}$ state suggests that the rf-field can induce transitions between collective states, so the F\"orster resonance occurs irrespective of the possibility of tuning it by the dc field alone. The dc field, however, should be applied to increase the efficiency, as will be discussed below.

Figures~2(b)-2(d) present the experimental records of the F\"orster resonance ${\rm Rb}(39P_{3/2} )+{\rm Rb}(39P_{3/2} )\to {\rm Rb}(39S_{1/2} )+{\rm Rb}(40S_{1/2} )$ in the rf-field at frequencies 90, 95, and 100 MHz, correspondingly. The rf-induced F\"orster resonance is clearly seen. Its position depends on the rf-frequency, while its width and height depend on the number of atoms [8,21]. For \textit{N}=2 the width is 18 mV/cm, corresponding to 1.1 MHz. At higher frequencies it appears at higher dc electric field. The resonance is quite efficient, as its amplitude is comparable with that of the "true" resonance [see Fig.~1(c)] and at \textit{N}=5 it is close to the maximum possible value of 0.25 for a disordered atom ensemble [21]. Figure~2 thus evidences the possibility to tune vdW to DD interaction with high efficiency using the rf-field, which is resonant to a quasi-molecular transition instead of atomic transition. 

\begin{figure}
\includegraphics[scale=0.42]{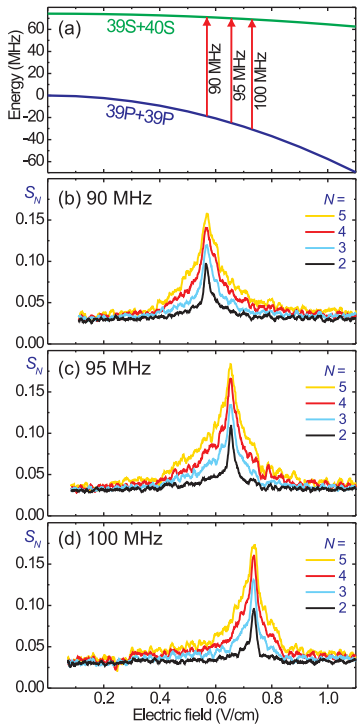}
\caption{\label{Fig2}(color online). (a) Energy levels of the initial 39\textit{P}+39\textit{P} and final 39\textit{S}+40\textit{S} collective states of two Rydberg atoms in the electric field. The dc electric field alone cannot tune a F\"orster resonance ${\rm Rb}(39P_{3/2} )+{\rm Rb}(39P_{3/2} )\to {\rm Rb}(39S)+{\rm Rb}(40S)$. The rf-photons couple these collective states and induce resonant dipole-dipole interaction. (b)-(d) RF-assisted F\"orster resonances for \textit{N}=2-5 detected Rydberg atoms at 100 mV amplitude and 90, 95, and 100~MHz rf-field frequency, respectively.}
\end{figure}

The physical interpretation of the rf-assisted F\"orster resonances was given in previous papers [15,18]. We will emphasize some important features in the two possible approaches. On the one hand, the rf-field induces transitions between quasi-molecular collective states, as shown in Figs.~1(a) and 2(a). Being absorbed or emitted by a quasi-molecule consisting of a few Rydberg atoms, the few rf-photons of frequency $\omega $ compensate for the energy defect $\Delta$ when it has values multiples of $\omega $.

On the other hand, rf-assisted F\"orster resonances can also be explained in terms of the Floquet sidebands induced by a periodic perturbation of the Rydberg energy levels by the rf electric field due to the Stark effect [15,18]. Following Ref.~[18], one should consider the Stark effect in a composite electric field consisting of dc and rf parts $F=F_{dc} +F_{rf} \; \cos (\omega t)$. The energy shift of a Rydberg level with nonzero quantum defect is quadratic and is given by its polarizability $E_{nL} =-\alpha _{nL} F^{2} /2$. This formula yields

\begin{equation} \label{Eq1} 
\begin{array}{l} {E_{nL} =-\dfrac{1}{2} \alpha _{nL} [F_{dc}^{2} +\dfrac{1}{2} F_{rf}^{2}+} \\ \\ {2F_{dc} F_{rf} \cos (\omega t)+\dfrac{1}{2} F_{rf}^{2}\cos (2 \omega t)]}. \end{array}  
\end{equation}

\noindent The term $F_{rf}^{2} /2$ in the brackets is responsible for the ac Stark shift of the Rydberg level in the rf-field [18]. The terms with $\cos (\omega t)$ and $\cos (2\omega t)$ drive the transitions between collective states as soon as the resonance condition $\Delta =m\omega $ is satisfied, with \textit{m} being an integer. This can be understood if we use a Floquet approach to find the eigenenergies of a Rydberg atom in dc+rf field [18]. It gives an infinite number of energy sidebands separated by $\omega $ and with relative amplitudes of the wave functions $a_{nL,m} $ described by the generalized Bessel functions

\begin{equation} \label{Eq2} 
a_{nL,m} =\sum _{k=-\infty }^{\infty }J_{m-2k} \left(\frac{\alpha _{nL} F_{dc} F_{rf} }{\omega } \right) J_{k} \left(\frac{\alpha _{nL} F_{rf}^{2} }{8 \omega } \right).    
\end{equation} 

The rf-assisted F\"orster resonances in this picture arise for the Floquet sidebands that satisfy the resonance condition $\Delta =m\omega $ and intersect at some particular values of the dc electric field. At $F_{dc} =0$ the odd sidebands disappear according to Eq.~\eqref{Eq2} since only $J_{0} (0)=1$ is nonzero, while the even sidebands are weak. Therefore, the rf-field alone hardly drives the transitions between quasi-molecular collective states with a quadratic Stark effect, so a dc field should also be present.

Figure~3(a) shows the energy levels of the initial 37\textit{P}+37\textit{P} and final 37\textit{S}+38\textit{S} collective states of two Rydberg atoms in electric field in the presence of the first Floquet sidebands at $\pm$15 MHz. The red (gray) circles indicate the intersections of the Floquet sidebands, corresponding to rf-assisted F\"orster resonances. These resonances are clearly seen in Fig.~3(b) on the experimental record at a 100 mV rf-amplitude. Figures~3(c) and 3(d) show the same for the F\"orster resonance on the 39\textit{P}$_{3/2}$ state at 95 MHz and 100 mV. There are much less Floquet intersections for this "inaccessible" F\"orster resonance, but a narrow second-order resonance at 1.55~V/cm is well seen along with a much stronger first-order resonance at 0.66~V/cm. The first-order resonance saturates and broadens as \textit{N} increases, while the second-order resonance is unsaturated and remains narrow for all \textit{N}.

\begin{figure}
\includegraphics[scale=0.42]{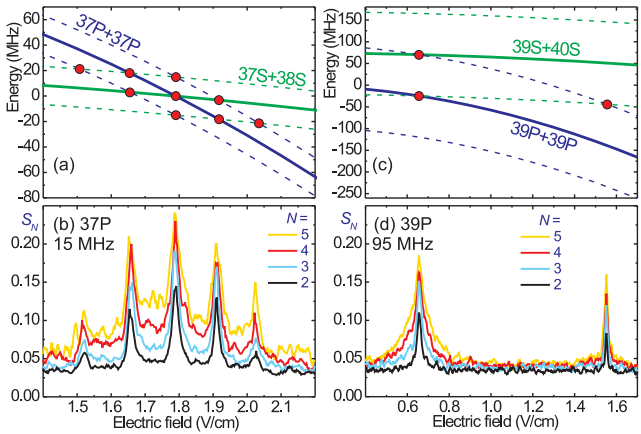}
\caption{\label{Fig3} (color online). (a) Energy levels of the initial 37\textit{P}+37\textit{P} and final 37\textit{S}+38\textit{S} collective states of two Rydberg atoms in the electric field in the presence of the first Floquet sidebands at $\pm$15~MHz. The red (gray) circles indicate the intersections of the Floquet sidebands corresponding to rf-assisted F\"orster resonances. (b) Experimental record of the rf-assisted F\"orster resonances at a 100~mV rf-amplitude for \textit{N}=2-5 detected Rydberg atoms. (c)-(d) The same for the "inaccessible" F\"orster resonances on the 39\textit{P} state at 95~MHz and 100~mV. The first- and second-order resonances are observed.}
\end{figure}

Below we discuss the impact of our experimental observations on the studies and applications of long-range interactions between Rydberg atoms. Radio-frequency-assisted "inaccessible" F\"orster resonances provide a way to tune vdW to resonant DD interactions and increase the interaction strength at long distances. The transition from vdW to DD can be analyzed by the formula describing the energy shift $\delta E_{PP}$ of the collective \textit{nP}+\textit{nP} state at the DD matrix element \textit{V} from \textit{nP}+\textit{nP} state to \textit{nS}+(\textit{n}+1)\textit{S} state and F\"orster detuning $\Delta $:

\begin{equation} \label{Eq3} 
\delta E_{PP} =\pm \left(\sqrt{\frac{\Delta ^{2} }{4} +2V^{2} } -\frac{\vert\Delta \vert}{2} \right).        
\end{equation} 

\noindent Here the sign is positive if the \textit{nP+nP} state lies above the \textit{nS}+(\textit{n}+1)\textit{S} state (this is the case for $n\le 38$), and vice versa. At $\Delta =0$ the interaction is purely DD, while at large detuning it is the vdW with energy shift $\pm 2V^{2} /\Delta =C_{6} /R^{6} $, where $C_{6} $ is vdW coefficient and \textit{R} is interatomic distance. The resonant rf-field with $\omega =\Delta / m$ compensates for $\Delta $ and tunes vdW to DD as if the $\Delta =0$ condition is satisfied.

This can be particularly useful for enhancing the dipole blockade effect in mesoscopic Rydberg ensembles [24]. For example, in a zero electric field the orientation-averaged energy of the vdW interaction between Rb(70\textit{P}$_{3/2}$) atoms at $\Delta /(2\pi )$=213~MHz and distance~10 $\mu $m is estimated to be 1.5~MHz, using $C_{6} \approx -1500$~GHz/$\mu$m$^6$ from Ref.~[25]. By applying $\sim$0.1~V/cm dc electric field and $\sim$0.05~V/cm rf-field at 250-280 MHz, it can be tuned to DD interaction with energy corresponding to 18~MHz, an order of magnitude larger than the vdW energy. For Rb \textit{nP}$_{3/2}$ states with \textit{n}=40-100 the required rf-frequencies lie in the 100-325~MHz range, and for \textit{nS} states with \textit{n}=70-120 they are in the 140-700~MHz range. These are reasonably low rf-frequencies, which can also be found in other alkali-metal atoms [26]. Radio-frequency-assisted F\"orster resonances thus significantly extend the range of the Rydberg states suitable for long-range resonant DD interaction. 

\begin{figure}
\includegraphics[scale=0.4]{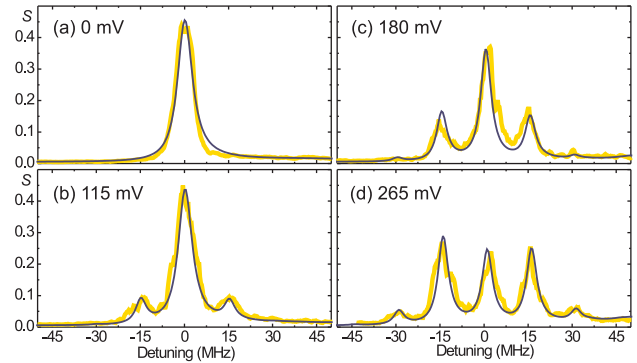}
\caption{\label{Fig4} (color online). Floquet sidebands observed at three-photon laser excitation of the 37\textit{P}$_{3/2}$ state in dc electric field of 1.8~V/cm with an admixed rf field at frequency 15~MHz for various rf-amplitudes: (a) 0~mV, (b) 115~mV, (c) 180~mV, (d) 265~mV. The yellow (gray) curves are experiment, and blue (dark) curves are theory calculated with our four-level density-matrix model [22]. }
\end{figure}

In our present experiment we first excited Rydberg atoms without the rf-field and then switched it on adiabatically [Fig.~1(b)]. In the experiments on dipole blockade, however, one needs the rf-field to be applied together with the laser excitation. Therefore, the laser radiation should be tuned in resonance not with the unperturbed Rydberg level, but with one of its Floquet sidebands that satisfies the F\"orster resonance condition $\Delta =m\omega $. We have observed these Floquet sidebands experimentally for the 37\textit{P}$_{3/2}$ state at $\omega/(2\pi)=$15 MHz and various rf-amplitudes, as shown in Fig.~4. The yellow (gray) curves are experimental, and blue (dark) curves are theoretical ones calculated with our four-level density-matrix model [22]. The sideband amplitudes are similar to those observed in Fig.~1 for the Stark switched excitation. Weaker Floquet sidebands have also been reported at Rydberg excitation in a Rb vapor cell [27].

To conclude, we have observed and studied the highly resolved rf-assisted F\"orster resonances between a few cold Rydberg atoms in a small laser excitation volume. These resonances correspond to single- and multiphoton rf-transitions between many-body collective states of a Rydberg quasi-molecule or to intersections of the Floquet sidebands of Rydberg levels appearing in the rf-field. We have shown that  they can be induced both for the "accessible" F\"orster resonances, which are tuned by the dc field, and for those which cannot be tuned and are "inaccessible". The van der Waals interaction of almost arbitrary high Rydberg states can thus be efficiently tuned to a resonant dipole-dipole interaction using the rf-field with frequencies below 1 GHz. This enhances the interaction strength and distance and can give rise to a much stronger dipole blockade effect.   

This work was supported by RFBR (Grants No.~13-02-00283 and No.~14-02-00680), by the Russian Academy of Sciences, by the EU FP7 IRSES Project "COLIMA", and by the Russian Quantum Center.


\begin{thebibliography}{10}

\bibitem{1} M.~Saffman, T.~G.~Walker, and K.~M$\o$lmer, Rev. Mod. Phys. \textbf{82}, 2313 (2010).

\bibitem{2} T.~Keating, K.~Goya, Y.~-Y.~Jau, G.~W.~Biedermann, A.~J.~Landahl, and I.~H.~Deutsch, Phys. Rev. A \textbf{87}, 052314 (2013).

\bibitem{3}  F.~Cinti, P.~Jain, M.~Boninsegni, A.~Micheli, P.~Zoller, and G.~Pupillo, Phys. Rev. Lett.\textbf{ 105}, 135301 (2010).

\bibitem{4}  J.~Honer, R.~L\"ow, H.~Weimer, T.~Pfau, and H.~P.~B\"uchler, Phys. Rev. Lett. \textbf{107}, 093601 (2011).

\bibitem{5} K.~A.~Safinya, J.~F.~Delpech, F.~Gounand, W.~Sandner, and T.~F.~Gallagher, Phys. Rev. Lett. \textbf{47}, 405 (1981).

\bibitem{6} A.~L.~de~Oliveira, M.~W.~Mancini, V.~S.~Bagnato, and L.~G.~Marcassa, Phys. Rev. Lett. \textbf{90}, 143002 (2003).

\bibitem{7} S.~Westermann, T.~Amthor, A.~L.~de~Oliveira, J.~Deiglmayr, M.~Reetz-Lamour, and M.~Weidem\"uller, Eur. Phys. J. D \textbf{40}, 37 (2006).

\bibitem{8}  I.~I.~Ryabtsev, D.~B.~Tretyakov, I.~I.~Beterov, and V.~M.~Entin, Phys. Rev. Lett. \textbf{104}, 073003 (2010).

\bibitem{9} P.~Bohlouli-Zanjani, J.~A.~Petrus, and J.~D.~D.~Martin, Phys. Rev. Lett. \textbf{98}, 203005 (2007); J.~A.~Petrus, P.~Bohlouli-Zanjani, and J.~D.~D.~Martin, J. Phys. B \textbf{41}, 245001 (2008).

\bibitem{10}  A.~Reinhard, K.~C.~Younge, and G.~Raithel, Phys. Rev. A \textbf{78}, 060702(R) (2008).

\bibitem{11}  J.~Nipper, J.~B.~Balewski, A.~T.~Krupp, B.~Butscher, R.~L\"ow, and T.~Pfau, Phys. Rev. Lett. \textbf{108}, 113001 (2012).

\bibitem{12} K.~Afrousheh, P.~Bohlouli-Zanjani, D.~Vagale, A.~Mugford, M.~Fedorov, and J.~D.~D.~Martin, Phys. Rev. Lett. \textbf{93}, 233001 (2004). 

\bibitem{13}  M.~Tanasittikosol, J.~D.~Pritchard, D.~Maxwell, A.~Gauguet, K.~J.~Weatherill, R.~M.~Potvliege, and C.~S.~Adams, J. Phys. B \textbf{44}, 184020 (2011).

\bibitem{14} E.~Brekke, J.~O.~Day, and T.~G.~Walker, Phys. Rev. A \textbf{86}, 033406 (2012).

\bibitem{15}  P.~Pillet, R.~Kachru, N.~H.~Tran, W.~W.~Smith, and T.~F.~Gallagher, Phys. Rev. A \textbf{36}, 1132 (1987).

\bibitem{16}  P.~Pillet, D.~Comparat, M.~Muldrich, T.~Vogt, N.~Zahzam, V.~M.~Akulin, T.~F.~Gallagher, W.~Li, P.~Tanner, M.~W.~Noel, and I.~Mourachko, in \textit{Decoherence, Entanglement and Information Protection in Complex Quantum Systems}, edited by V.~M.~Akulin \textit{et al.}, Springer, 2005, p.411.

\bibitem{17}  A.~Tauschinsky, C.~S.~E.~van~Ditzhuijzen, L.~D.~Noordam, and H.~B.~van~Linden van~den~Heuvell, Phys. Rev. A \textbf{78}, 063409 (2008).

\bibitem{18}  C.~S.~E.~van~Ditzhuijzen, A.~Tauschinsky, and H.~B.~van~Linden van~den~Heuvell, Phys. Rev. A \textbf{80}, 063407 (2009).

\bibitem{19}  Y.~Yu, H.~Park, and T.~F.~Gallagher, Phys. Rev. Lett. \textbf{111}, 173001 (2013).

\bibitem{20}  D.~B.~Tretyakov, I.~I.~Beterov, V.~M.~Entin, I.~I.~Ryabtsev, and P.~L.~Chapovsky, J. Exper. Theor. Phys. \textbf{108}, 374 (2009).

\bibitem{21}  I.~I.~Ryabtsev, D.~B.~Tretyakov, I.~I.~Beterov, V.~M.~Entin, and E.~A.~Yakshina,  Phys. Rev. A \textbf{82}, 053409 (2010).

\bibitem{22}  V.~M.~Entin, E.~A.~Yakshina, D.~B.~Tretyakov, I.~I.~Beterov, and I.~I.~Ryabtsev, J. Exper. Theor. Phys. \textbf{116}, 721 (2013).

\bibitem{23} I.~I.~Ryabtsev, D.~B.~Tretyakov, I.~I.~Beterov, and V.~M.~Entin, Phys. Rev. A \textbf{76}, 012722 (2007); Erratum: Phys. Rev. A \textbf{76}, 049902(E) (2007).

\bibitem{24}  M.~D.~Lukin, M.~Fleischhauer, R.~Cote, L.~M.~Duan, D.~Jaksch, J.~I.~Cirac, and P.~Zoller, Phys. Rev. Lett. \textbf{87}, 037901 (2001); D.~Comparat and P.~Pillet, J. Opt. Soc. Am. B \textbf{27}, A208 (2010).

\bibitem{25}  T.~G.~Walker and M.~Saffman, Phys. Rev. A \textbf{77}, 032723 (2008).

\bibitem{26}   J.~H.~Gurian, P.~Cheinet, P.~Huillery, A.~Fioretti, J.~Zhao, P.~L.~Gould, D.~Comparat, and P.~Pillet, Phys. Rev. Lett. \textbf{108}, 023005 (2012).

\bibitem{27}  M.~G.~Bason, M.~Tanasittikosol, A.~Sargsyan, A.~K.~Mohapatra, D.~Sarkisyan, R.~M.~Potvliege, and C.~S.~Adams, New J. Phys. \textbf{12}, 065015 (2010).

\end{thebibliography}
\end{document}